\documentclass[12pt]{article}
\usepackage[top=22truemm,bottom=22truemm,left=20truemm,right=20truemm]{geometry}
\usepackage{tikz}
\usepackage{amssymb}
\usepackage{braket}
\usepackage{amsmath}
\usepackage{lipsum}
\usepackage{csquotes}
\usepackage{braket}
\usepackage{enumitem}
\usepackage{color}
\usepackage{hyperref}
\usepackage{pgfplots}
\usepackage{tikz}
\usepackage{graphicx}

\newcommand{\be}{\begin{equation}}
\newcommand{\ee}{\end{equation}}
\newcommand{\ba}{\begin{eqnarray}}
\newcommand{\ea}{\end{eqnarray}}

\makeatletter

\@addtoreset{equation}{section}

\newtheorem{kdefi}{Definition}[section]
\newtheorem{klemma}[kdefi]{Lemma}

\newtheorem{kfact}[kdefi]{Fact}

\makeatother

\makeatletter

\begin{document} 

\title{Qubit Transport Model for Unitary Black Hole Evaporation without Firewalls 
\thanks{Alberta Thy 25-16, arXiv: 1607.04642[hep-th]} }
\author{Kento Osuga\thanks{Email address: osuga@ualberta.ca}  \;and\; Don N. Page
\thanks{Email address: profdonpage@gmail.com}\\[.5cm] Theoretical Physics Institute \\ Department of Physics\\ 4-181 CCIS\\ University of Alberta \\ Edmonton, Alberta T6G 2E1\\ Canada}

\date{January 10, 2018}
\maketitle

\begin{abstract}
We give an explicit toy qubit transport model for transferring information from the gravitational field of a black hole to the Hawking radiation by a continuous unitary transformation of the outgoing radiation and the black hole gravitational field. The model has no firewalls or other drama at the event horizon, and it avoids a counterargument that has been raised for subsystem transfer models as resolutions of the firewall paradox. Furthermore, it fits the set of six physical constraints that Giddings has proposed for models of black hole evaporation. It does utilize nonlocal qubits for the gravitational field but assumes that the radiation interacts locally with these nonlocal qubits, so in some sense the nonlocality is confined to the gravitational sector. Although the qubit model is too crude to be quantitatively correct for the detailed spectrum of Hawking radiation, it fits qualitatively with what is expected.
\end{abstract}
\newpage
\section{Introduction}

The black hole information puzzle is the puzzle of whether black hole formation and evaporation is unitary, and debate on this issue has continued for more than 36 years \cite{Page:1993up, Giddings:2006sj, Mathur:2008wi}, since Hawking radiation was discovered \cite{Hawking:1974sw}. Hawking originally used local quantum field theory in the semiclassical spacetime background of an evaporating black hole to deduce \cite{Hawking:1976ra} that part of the information about the initial quantum state would be destroyed or leave our Universe at the singularity or quantum gravity region at or near the centre of the black hole, so that what remained outside after the black hole evaporated would not be given by unitary evolution from the initial state.  

However, this approach does not fully apply quantum theory to the gravitational field itself, so it was objected that the information-loss conclusion drawn from it might not apply in quantum gravity \cite{Page:1979tc}. Maldacena's AdS/CFT conjecture \cite{Maldacena:1997re} has perhaps provided the greatest impetus for the view that quantum gravity should be unitary within our Universe and give no loss of information.

If one believes in local quantum field theory outside a black hole and also that one would not experience extreme harmful conditions (`drama') immediately upon falling into any black hole sufficiently large that the curvature at the surface would not be expected to be dangerous, then recent papers by Almheiri, Marolf, Polchinski, and Sully (AMPS) \cite{Almheiri:2012rt}, and by them and Stanford (AMPSS) \cite{Almheiri:2013hfa}, give a new challenge to unitarity, as they argued that unitarity, locality, and no drama are mutually inconsistent.

It seems to us that locality is the most dubious of these three assumptions.  Nevertheless, locality seems to be such a good approximation experimentally that we would like a much better understanding of how its violation in quantum gravity might be able to preserve unitarity and yet not lead to the drama of firewalls or to violations of locality so strong that they would be inconsistent with our observations.  Giddings (occasionally with collaborators) has perhaps done the most to investigate unitary nonlocal models for quantum gravity \cite{Giddings:2006sj, Giddings:2006be, Giddings:2007ie, Giddings:2007pj, Giddings:2009ae, Giddings:2011ks, Giddings:2012bm, Giddings:2012dh, Giddings:2012gc, Giddings:2013kcj, Giddings:2013jra, Giddings:2013noa, Giddings:2014nla, Giddings:2014ova, Giddings:2015uzr, Donnelly:2016rvo, Giddings:2017mym, Donnelly:2017jcd}.  For other black hole qubit models, see \cite{Terno:2005ff, Levay:2006pt, Levay:2007nm, Duff:2008eei, Levay:2008mi, Borsten:2008wd, Rubens:2009zz, Levay:2010ua, Duff:2010zz, Duff:2012nd, Borsten:2011is, Levay:2011bq, Avery:2011nb, Dvali:2011aa, Borsten:2012sga, Borsten:2012fx, Dvali:2012en, Duff:2013xna, Levay:2013epa, Verlinde:2013vja, Borsten:2013vea, Duff:2013rma, Borsten:2013uma, Dvali:2013lva, Prudencio:2014ypa, Pramodh:2014jha, Chatwin-Davies:2015hna, Dai:2015dqt, Belhaj:2016yyq, Belhaj:2016yfo}.

Here we present a qubit toy model for how a black hole might evaporate unitarily and without firewalls, but with nonlocal gravitational degrees of freedom. We model radiation modes emitted by a black hole as localized qubits that interact locally with these nonlocal gravitational degrees of freedom.  Similar models were first investigated by Giddings in his previously referred papers, particularly in \cite{Giddings:2011ks,Giddings:2012bm,Giddings:2012dh}. Nomura and his colleagues also have a model \cite{Nomura:2014woa,Nomura:2014voa,Nomura:2016qum} with some similarities to ours. In this way we can go from modes near the horizon that to an infalling observer appear to be close to a vacuum state (and hence without a firewall), and yet the modes that propagate outward can pick up information from the nonlocal gravitational field they pass through so that they transfer that information out from the black hole.

\section{Qualitative Description of Our Qubit Model}

Using Planck units in which $\hbar = c = G = k_\mathrm{Boltzmann} = 1$, a black hole that forms of area $A$ and Bekenstein-Hawking entropy $S_\mathrm{BH} = A/4$ may be considered to have $e^{S_\mathrm{BH}} = 2^{S_\mathrm{BH}/(\ln{2})}$ orthonormal states, which is the same number as the number of orthonormal states of $n = S_\mathrm{BH}/(\ln{2}) = A/(4\ln{2})$ qubits if this is an integer, which for simplicity we shall assume.  We shall take the state of these $n$ qubits as being the state of the gravitational field of the black hole.  We assume that this state is rapidly scrambled by highly complex unitary transformations, so that generically a black hole formed by collapse, even if it is initially in a pure state, will have these $n$ qubits highly entangled with each other.

However, in our model we shall assume that there are an additional $n$ qubits of outgoing radiation modes just outside the horizon, and a third set of $n$ qubits of outgoing but infalling radiation modes just inside the horizon.  We shall assume that these two sets of qubits have a unique pairing (as partner modes in the beginning of the Hawking radiation) and further that each pair is in the singlet Bell state that we shall take to represent the vacuum state as seen by an infalling observer, so that all of these $2n$ qubits of radiation modes near the black hole horizon are in the vacuum pure state and hence give no contribution to the Bekenstein-Hawking entropy $S_\mathrm{BH} = n\ln{2}$. We thus explicitly assume that the infalling observer sees only the vacuum and no firewall in crossing the event horizon.  See \cite{Page:2013mqa} for one argument for justifying this assumption.

Now we assume that the Hawking emission of one mode corresponds to one of the $n$ outgoing radiation modes from just outside the horizon propagating to radial infinity.  However, the new assumption of this model is that the radiation qubit that propagates outward interacts (locally) with one of the $n$ nonlocal qubits representing the black hole gravitational field, in just such a way that when the mode gets to infinity, the quantum state of that radiation qubit is interchanged with the quantum state of the corresponding black hole gravitational field qubit.  This is a purely unitary transformation, not leading to any loss of information.

Assume for simplicity that the black hole forms in a pure state that becomes highly scrambled by a unitary transformation.  Therefore, as an early outgoing radiation qubit propagates out to become part of the Hawking radiation, when it interchanges its state with that of the corresponding gravitational field qubit, it will become nearly maximally entangled with the black hole state and will have von Neumann entropy very nearly $\ln{2}$, the maximum for a qubit.  So the early Hawking radiation qubits will each have nearly the maximum entropy allowed, and there will be very little entanglement between the early radiation qubits themselves.

Meanwhile, the black hole qubit corresponding to each outgoing radiation qubit will have taken on the state that the outgoing radiation qubit had when it was just outside the horizon and hence be in the unique singlet Bell state with the infalling radiation qubit just inside the horizon that was originally paired with the outgoing qubit.  This vacuum singlet Bell state can then be omitted from the analysis without any loss of information.  In this way we can model the reduction in the size of the black hole as it evaporates by the reduction of the number of black hole qubits. We might say that each such vacuum Bell pair falls into the singularity, but what hits the singularity in this model is a unique quantum state, similar to the proposal of Horowitz and Maldacena \cite{Horowitz:2003he}.

Therefore, if we start with $n$ black hole gravitational field qubits, $n$ outgoing radiation qubits just outside the horizon, and $n$ infalling radiation qubits just inside the horizon, after the emission of $n_r$ outgoing radiation qubits, $n_r$ of the infalling radiation qubits will have combined into a unique quantum state with the $n_r$ black hole qubits that were originally interacting with the $n_r$ outgoing radiation qubits that escaped, so that we can ignore them as what we might regard as merely vacuum fluctuations.  This leaves $n-n_r$ pairs of outgoing radiation qubits just outside the horizon and infalling qubits just inside the horizon (each pair being in the singlet Bell state), and $n-n_r$ black hole gravitational field qubits.

Eventually the number of Hawking radiation qubits, $n_r$, exceeds the number of black hole qubits remaining, $n-n_r$, when $n_r > n/2$, and the black hole becomes `old.' At this stage, the remaining black hole qubits all become nearly maximally entangled with the Hawking radiation qubits, so that the von Neumann entropy of the black hole becomes very nearly $(n-n_r)\ln{2}$, which we shall assume is very nearly $A/4$ at that time.  Since the whole system is assumed to be in a pure state, and since we have assumed unitary evolution throughout, the von Neumann entropy of the Hawking radiation at this late stage is also very nearly $(n-n_r)\ln{2}$, but now this is less than the maximum value, which is $n_r\ln{2}$. Thus each of the $n_r$ Hawking radiation qubits can no longer be maximally entangled with the remaining $n-n_r$ black hole qubits, and significant entanglement begins to develop between the Hawking radiation qubits themselves.  Nevertheless, for any collection of $n' < n/2$ qubits of the Hawking radiation, the von Neumann entropy of that collection is expected \cite{Page:1993df, Page:1993wv} to be very nearly $n'\ln{2}$, so one would still find negligible quantum correlations between any collection of $n'$ Hawking radiation qubits.

Finally, when all $n$ of the original outgoing radiation qubits have left the black hole and propagated to infinity to become Hawking radiation qubits, there are no qubits left for the black hole; hence it has completely evaporated away. The $n$ Hawking radiation qubits now form a pure state, just as the original quantum state that formed the black hole was assumed to be. Of course, the unitary scrambling transformation of the black hole qubits means that the pure state of the final Hawking radiation can look quite different from the initial state that formed the black hole, but the two are related by a unitary transformation.

The net effect is that the emission of one outgoing radiation qubit gives the transfer of the information in one black hole qubit to one Hawking radiation qubit.  But rather than simply saying that this transfer is nonlocal, from the inside of the black hole to the outside, we are saying that the black hole qubit itself is always nonlocal, and that the outgoing radiation qubit picks up the information in the black hole qubit locally, as it travels outward through the nonlocal gravitational field of the black hole.  Therefore, in this picture in which we have separated the quantum field theory qubits of the radiation from the black hole qubits of the gravitational field, we do not need to require any nonlocality for the quantum field theory modes, but only for the gravitational field.  In this way the nonlocality of quantum gravity might not have much observable effect on experiments in the laboratory focussing mainly on local quantum field theory modes.

\section{Mathematics of Qubit Transport}

Before the black hole forms, we assume that we have a Hilbert space of dimension $2^n$ in which each state collapses to form a black hole whose gravitational field can be represented by $n$ nonlocal qubits.  We assume that we have a pure initial state represented by the set of $2^n$ amplitudes $A_{q_1q_2\ldots q_n}$, where for each $i$ running from 1 to $n$, the corresponding $q_i$ can be 0 or 1, representing the two basis states of the $i$th qubit.  Once the black hole forms, without changing the Hilbert space dimension, we can augment this Hilbert space by taking its tensor product with a 1-dimensional Hilbert space for the vacuum state of $n$ infalling and $n$ outgoing radiation modes just inside and just outside the event horizon.  We shall assume that this vacuum state is the tensor product of vacuum states for each pair of modes, with each pair being in the singlet Bell state that we shall take to represent the vacuum for that pair of modes.

That is, once the black hole forms, we assume that we have $n$ nonlocal qubits for the gravitational field of the black hole, labeled by $a_i$, where $i$ runs from 1 to $n$, $n$ localized qubits for the infalling radiation modes just inside the horizon, labeled by $b_i$, and $n$ localized qubits for the outgoing radiation modes just outside the horizon, labeled by $c_i$.  Suppose that each qubit has basis states $\ket{0}$ and $\ket{1}$, where subscripts (either $a_i$, $b_i$, or $c_i$) will label which of the $3n$ qubits one is considering.  We assume that each pair of infalling and outgoing radiation qubits is in the vacuum singlet Bell state
\be
\ket{B}_{b_i c_i} = \frac{1}{\sqrt{2}}\Bigl(\ket{0}_{b_i}\ket{1}_{c_i} -\ket{1}_{b_i}\ket{0}_{c_i}\Bigr).
\label{Bell}
\ee
Initially the quantum state of the black hole gravitational field and radiation modes is
\be
\ket{\Psi_0}=\sum_{q_1=0}^1\sum_{q_2=0}^1\cdots\sum_{q_n=0}^1 A_{q_1q_2\ldots q_n}\prod_{i=1}^n\ket{q_i}_{a_i}\prod_{i=1}^n\ket{B}_{b_ic_i},
\label{initial state}
\ee
where the $A_{q_1q_2\ldots q_n}$ are the amplitudes for the $2^n$ product basis states for the black hole gravitational field.  Note that the entire quantum state is the product of a state of all the black hole gravitational qubits and a single pure vacuum state for the radiation modes.

During the emission of the $i$th radiation mode to become a mode of Hawking radiation at radial infinity, the basis state for the subsystem of the $i$th black hole, infalling radiation, and outgoing radiation qubits changes as
\be
\ket{q_i}_{a_i}\ket{B}_{b_ic_i} \mapsto
-\ket{B}_{a_ib_i}\ket{q_i}_{c_i},
\label{transfer}
\ee
where $\ket{B}_{a_ib_i}$ is the analogue of $\ket{B}_{b_ic_i}$ given by Eq.\ (\ref{Bell}) with $b_i$ replaced by $a_i$ and $c_i$ replaced by $b_i$. As is obvious from the expressions on the right hand sides, this just interchanges the state of the $i$th black hole qubit with the state of the $i$th outgoing radiation qubit.

If $P_{a_ic_i} = \ket{B}_{a_ic_i}\bra{B}_{a_ic_i}$ multiplied by the identity operator in the $b_i$ subspace, then for $\theta = \pi$ the continuous sequence of unitary transformations
\be
U(\theta)=\exp\Bigl(-i\theta P_{a_ic_i}\Bigr)={\rm I}+(e^{-i\theta}-1)P_{a_ic_i}
\label{Unitary operator for qubit transfer}
\ee
becomes $U(\pi) = {\rm I}-2P_{a_ic_i}$, which gives the unitary transformation \eqref{transfer}, interchanging the states of the $i$th black hole qubit with the state of the outgoing radiation qubit.

We might suppose that as the radiation qubit moves outward, the $\theta$ parameter of the unitary transformation is a function of the radius $r$ that changes from 0 at the horizon to $\pi$ at radial infinity.  For example, one could take $\theta = \pi(1 - K/K_h)$, where $K$ is some curvature invariant (such as the Kretschmann invariant, $K =   R^{\mu\nu\rho\sigma}R_{\mu\nu\rho\sigma}$) that decreases monotonically from some positive value at the horizon (where its value is $K_h$) to zero at infinity.

We now assume that after the emission of the $i$th mode, the vacuum Bell state of the $i$th black hole qubit and the $i$th infalling radiation qubit can be dropped from the analysis, so that one only has the Hawking radiation qubit remaining for that $i$. Then the state of the subsystem for that $i$ goes from $-\ket{B}_{a_ib_i}\ket{q_i}_{c_i}$ given by Eq.\ (\ref{transfer}) to simply $\ket{q_i}_{c_i}$ for the qubit representing the Hawking radiation mode.  Therefore, after all of the $n$ outgoing radiation modes propagate out to infinity while interacting with the black hole gravitational field, and after all the Bell vacua left inside the black hole are omitted, one is left with no black hole and the Hawking radiation in the final pure state
\be
\ket{\Psi_1}=\sum_{q_1=0}^1\sum_{q_2=0}^1\cdots\sum_{q_n=0}^1 A_{q_1q_2\ldots q_n}\prod_{i=1}^n\ket{q_i}_{c_i}.
\label{final state}
\ee

As a note we require that nonlocal gravitational qubits $a_i$ do not create firewalls by themselves. That is, even though the vacuum states on the horizon $b_i,c_i$ are in the range of nonlocal effects, they remain to be constrained in the singlet state unless systems $c_i$ are propagating away to infinity as Hawking radiation by Eq.\ \eqref{Unitary operator for qubit transfer}. This is consistent with the above assumption that the parameter $\theta$ in Eq.\ \eqref{Unitary operator for qubit transfer} is a function of the radius $r$. Conversely, it seems plausible to assume that any incoming mode gradually \emph{drops off} some of its information during propagation through this nonlocal gravitational field.

\subsection{Mining Issue}
AMPSS \cite{Almheiri:2013hfa}, whose Eq.\ (3.3) is essentially the same as our \eqref{transfer}, raised the following issue with subsystem transfer models as resolutions of the firewall paradox. Suppose there exists an ideal mining equipment that can approach arbitrarily close to the horizon without falling into it, and then the equipment interacts with one of systems $c_i$ just outside the horizon. Note that this can be done without any exchange of energy due to the infinite redshift, and it is assumed that there is no entangling either. For example, the mining equipment can unitarily acts on the system $c_i$ as
\begin{eqnarray}
U_{\text{mine}}&:&\ket{0}_{c_i}\mapsto e^{i\phi}\ket{0}_{c_i},\;\;\;\ket{1}_{c_i}\mapsto e^{-i\phi}\ket{1}_{c_i}.\\
U_{\text{mine}}&:&\ket{B}_{b_ic_i}\mapsto\frac{\cos\phi}{\sqrt{2}}\Bigl(\ket{0}_{b_i}\ket{1}_{c_i} -\ket{1}_{b_i}\ket{0}_{c_i}\Bigr)+\frac{i\sin\phi}{\sqrt{2}}\Bigl(\ket{0}_{b_i}\ket{1}_{c_i} +\ket{1}_{b_i}\ket{0}_{c_i}\Bigr).\label{mine}
\end{eqnarray}
Thus the system on the horizon has one bit of information after this mining process and is thus no longer in the vacuum state.

First of all, it seems implausible that such an ideal equipment can be physically realistic. Since the equipment is accelerating in order to stay outside the horizon without falling into the black hole, it has an Unruh temperature that becomes very high near the horizon. Then the equipment and the modes it interacts with, $c_i$ in this case, should strongly couple and would be expected to be approximately in a thermal state. As a consequence it seems plausible that energy must be transferred between the mining equipment and the modes $c_i$.

Also, notice that the AMPSS mining argument does not take nonlocality into account. That is, the mining equipment would interact with the nonlocal gravitational degrees of freedom even if it could avoid the objection of the previous paragraph. As discussed previously, interactions with nonlocal gravitational degrees of freedom transfer part of the quantum information of the mining system into the gravitational degrees of freedom as the equipment approaches to the horizon. We can think of this transferred part as now being a part of the temporarily enlarged nonlocal gravitational degrees of freedom when the equipment is very near to the horizon. Then in this picture the mining equipment can still produce the phase change Eq.\ \eqref{mine} on the system just outside the horizon, but this excitation will be eventually absorbed into the nonlocal gravitational degrees of freedom. This absorption is possible regardless of how old the black hole is, because the nonlocal degrees of freedom are temporarily enlarged by the partially transferred degrees of freedom of the mining equipment. In summary, the AMPSS mining argument is not problematic for our model.

\section{Giddings' Physical Conditions}
Giddings \cite{Giddings:2012bm} has proposed a list of physical constraints on models of black hole evaporation.  We shall write each constraint in italics below and then follow that with comments on how our qubit model can satisfy the proposed constraint.

(i) \emph{Evolution is unitary.}  Our model explicitly assumes unitary evolution.

(ii) \emph{Energy is conserved.}  Our model is consistent with a conserved energy given by the asymptotic behavior of the gravitational field.  The unitary transformation $U(\theta(r))$ during the propagation of each radiation qubit can be written in terms of a radially dependent Hamiltonian without any explicit time dependence, so there is nothing in our model that violates energy conservation.

(iii) \emph{The evolution should appear innocuous to an infalling observer crossing the horizon; in this sense the horizon is preserved.}  We explicitly assume that the radiation modes are in their vacuum states when they are near the horizon, so there is no firewall or other drama there.

(iv) \emph{Information escapes the black hole at a rate $dS/dt\sim1/R$.}  Although we did not discuss the temporal rates above, if one radiation qubit propagates out through some fiducial radius, such as $r = 3M$, during a time period comparable to the black hole radius $R$, since during the early radiation each qubit carries an entropy very nearly $\ln{2}$, indeed one would have $dS/dt\sim1/R$.

(v) \emph{The coarse-grained features of the outgoing radiation are still well-approximated as thermal.}  Because of the scrambling of the black hole qubits so that each one is very nearly in a maximally mixed state, when the information is transferred from the black hole qubits to the Hawking radiation qubits, each one of these will also be very nearly in a maximally mixed state, which in the simplified toy model represents thermal radiation.  Furthermore, one would expect that any collection of $n' < n/2$ qubits of the Hawking radiation also to be nearly maximally mixed, so all the coarse-grained features of the radiation would be well-approximated as thermal.

(vi) \emph{Evolution of a system ${\cal H}_A\otimes{\cal H}_B$ saturates the subadditivity inequality $S_A+S_B \geq S_{AB}$.}  Here it is assumed that $A$ and $B$ are subsystems of $n_A$ and $n_B$ qubits respectively of the black hole gravitational field and of the Hawking radiation, not including any of the infalling and outgoing radiation qubits when they are near the horizon.  Then for $n_A + n_B < n/2$, $A$, $B$, and $AB$ are all nearly maximally mixed, so $S_A \approx n_A\ln{2}$, $S_B \approx n_B\ln{2}$, and $S_{AB} \approx (n_A+n_B)\ln{2}$, thus approximately saturating the subadditivity inequality.  (Of course, for any model in which the total state of $n$ qubits is pure and any collection of $n' < n/2$ qubits has nearly maximal entropy, $S \approx n'\ln{2}$, then if $n_A < n/2$, $n_B < n/2$, but $n_A + n_B > n/2$, then $S_A \approx n_A\ln{2}$ and $S_B \approx n_B\ln{2}$, but $S_{AB} \approx (n-n_A-n_B)\ln{2}$, so $S_A+S_B-S_{AB} \approx 2n_A+2n_B-n > 0$, so that the subadditivity inequality is generically not saturated in this case.)
  
\section{Conclusions}

We have given a toy qubit model for black hole evaporation that is unitary and does not have firewalls.  It does have nonlocal degrees of freedom for the black hole gravitational field, but the quantum field theory radiation modes interact purely locally with the gravitational field, so in some sense the nonlocality is confined to the gravitational sector.  The model has no mining issue and also satisfies all of the constraints that Giddings has proposed, though further details would need to be added to give the detailed spectrum of Hawking radiation. The model is in many ways {\it ad hoc}, such as in the details of the qubit transfer, so one would like a more realistic interaction of the radiation modes with the gravitational field than the simple model sketched here.  One would also like to extend the model to include possible ingoing radiation from outside the black hole.

\section*{Acknowledgments}
DNP acknowledges discussions with Beatrice Bonga, Fay Dowker, Jerome Gauntlett, Daniel Harlow, Adrian Kent, Donald Marolf, Jonathan Oppenheim, Subir Sachdev, and Vasudev Shyam at the Perimeter Institute, where an early version of this paper was completed. We also benefited from emails from Steven Avery, Giorgi Dvali, Steven Giddings, Yasunori Nomura, and Douglas Stanford. Revisions were made while using Giorgi Dvali's office during the hospitality of Matthew Kleban at the Center for Cosmology and Particle Physics of New York University.  This research was supported in part by the Natural Sciences and Engineering Research Council of Canada, and in part by Perimeter Institute for Theoretical Physics.  Research at Perimeter Institute is supported by the Government of Canada through the Department of Innovation, Science and Economic Development and by the Province of Ontario through the Ministry of Research, Innovation and Science.

\end{document}